\documentclass[a4paper,12pt]{article}
\usepackage{graphicx,amssymb,bm,latexsym}
\newcommand{\bea}{\begin{eqnarray}}
\newcommand{\ena}{\end{eqnarray}}
\newcommand{\vs}[1]{\vspace{#1 mm}}
\newcommand{\hs}[1]{\hspace{#1 mm}}
\renewcommand{\a}{\alpha}
\renewcommand{\b}{\beta}
\renewcommand{\c}{\gamma}
\renewcommand{\d}{\delta}
\newcommand{\e}{\epsilon}

\newcommand{\shalf}{\frac{1}{2}}
\newcommand{\pa}{\partial}
\newcommand{\td}{{\tilde d}}
\newcommand{\nn}{\nonumber\\}
\newcommand{\p}[1]{(\ref{#1})}

\begin{document}
\topmargin 0pt
\oddsidemargin 5mm

\begin{titlepage}

\begin{flushright}
OU-HET 471 \\
hep-th/0404082
\end{flushright}

\vs{10}
\begin{center}
{\Large\bf Complete Intersecting Non-Extreme $p$-Branes}
\vs{30}

{\large
Yan-Gang Miao\footnote{e-mail address:
 miao@het.phys.sci.osaka-u.ac.jp}
and
Nobuyoshi Ohta\footnote{e-mail address: ohta@phys.sci.osaka-u.ac.jp}}\\
\vspace{10mm}
{\em Department of Physics, Osaka University,
Toyonaka, Osaka 560-0043, Japan}

\end{center}

\vs{15}
\centerline{{\bf{Abstract}}}
\vs{5}

We give general intersecting brane solutions without assuming any
restriction on the metric in supergravity coupled to a dilaton and
antisymmetric tensor fields in arbitrary dimensions $D$.
The result is a general class of intersecting brane solutions which
interpolate the non-extreme solutions of type 1 and 2. We also discuss
the relation of our solutions to the known single brane solution.

\end{titlepage}
\newpage
\setcounter{page}{2}

Understanding classical solutions of supergravities in eleven and ten
dimensions is an important subject in the current particle physics.
These are the low-energy effective theories of string and M theories.
An important class of solutions in such theories are
the extended objects called branes~\cite{DGH}-\cite{DLP}, which have
played significant role in our study of nonperturbative effects in
strings and field theories realized on the branes.
In particular non-extreme solutions give rise to non-extreme black holes
and thus are very important in studying the properties of realistic
black holes. Various supersymmetric and non-extreme solutions, and their
intersections have been studied so far~\cite{DKL}-\cite{OPS}.

It has been known that there are two possible ways to construct non-extreme
solutions, classified as type 1 and 2 in ref.~\cite{LMP}. Type 1 has
the metric
\bea
ds^2= e^{2A} dx^2_{p+1} + e^{2B}(dr^2+r^2 d\Omega_{\td+1}^2),
\label{t1}
\ena
where the dimension of the space-time is given as $D=p+\td+3$ and
there is no restriction on the functions $A$ and $B$ except that they
are functions of $r$ only. The usual extreme solutions are obtained under
the condition~\cite{DKL}
\bea
(p+1)A+\td B=0,
\label{x}
\ena
which can be understood as `no-force' or BPS condition.
By type 1 non-extreme solutions, we mean that the restriction \p{x} is removed.

The metric for type 2 solutions is taken as
\bea
ds^2= e^{2A} (-f dt^2+ dx^2_p) + e^{2B}(f^{-1}dr^2+r^2 d\Omega_{\td+1}^2),
\label{t2}
\ena
with the restriction~\p{x}. Here the function $f$ gives the non-extreme
extension.

There have been many works on these two kinds of non-extreme solutions
separately~\cite{PO}-\cite{GLC}, but to the best of our knowledge
neither clarification of the connection of these solutions nor
attempt at interpolating these two classes of solutions have been made.
In view of the importance of both these solutions,
it is interesting to examine if there are more general solutions that
include both classes of solutions and hence interpolate these in the
particular limits of the parameters.
The purpose of this paper is to show that this is indeed possible by
deriving complete intersecting brane solutions without the restriction~\p{x}.
We also discuss their relations to other known solutions.

The method adopted here is a simple generalization of that developed
by one of the present authors some time ago~\cite{NO1} for the type 2
solutions. There the field equations were solved with a simplifying
ansatz which generalizes the condition~\p{x}. What we show here is that
it is in fact possible to solve
the field equations without this ansatz, and the result is a very general
class of solutions that involve additional integration constants, and
their appropriate choices give both the solutions of type 1 and 2.

Let us start with the general action for gravity coupled to a dilaton
$\phi$ and $m$ different $n_A$-form field strengths:
\bea
I = \frac{1}{16 \pi G_D} \int d^D x \sqrt{-g} \left[
R - \shalf (\pa \phi)^2 - \sum_{A=1}^m \frac{1}{2 n_A!} e^{a_A \phi}
F_{n_A}^2 \right].
\label{act}
\ena
This action describes the bosonic part of $D=11$ or $D=10$ supergravities;
we simply drop $\phi$ and put $a_A=0$ and $n_A=4$ for $D=11$, whereas we
set $a_A=-1$ for the NS-NS 3-form and $a_A=\shalf(5-n_A)$ for forms coming
from the R-R sector.\footnote{There may be Chern-Simons terms in the action,
but they are irrelevant in our following solutions.} To describe more
general supergravities in lower dimensions, we should include several scalars
as in ref.~\cite{PO}, but for simplicity we disregard this complication
in this paper.

From the action (\ref{act}), one derives the field equations
$$
R_{\mu\nu} = \shalf \pa_\mu \phi \pa_\nu \phi + \sum_{A} \frac{1}{2 n_A!}
e^{a_A \phi} \left[ n_A \left( F_{n_A}^2 \right)_{\mu\nu}
- \frac{n_A -1}{D-2} F_{n_A}^2 g_{\mu\nu} \right],
$$
$$
\Box \phi = \sum_{A} \frac{a_A}{2 n_A!} e^{a_A \phi} F_{n_A}^2,
$$
$$
\pa_{\mu_1} \left( \sqrt{- g} e^{a_A \phi} F^{\mu_1 \cdots \mu_{n_A}} \right)
= 0,
$$ \vs{-10}
\bea
\pa _{[\mu} F_{\mu_1 \cdots \mu_{n_A}]} = 0.
\label{fe}
\ena
The last equations are the Bianchi identities.

We take the following metric for our system:
\bea
ds_D^2 = -e^{2u_0} f dt^2 + \sum_{\a=1}^{p} e^{2 u_\a} dy_\a^2
+ e^{2B} \left[ f^{-1} dr^2 + r^2 d\Omega_{\td+1}^2 \right],
\label{met}
\ena
where $D=p+\td+3$, the coordinates $y_\a, (\a=1,\ldots, p)$ parametrize the
$p$-dimensional compact directions and the remaining coordinates of
the $D$-dimensional spacetime are the radius $r$ and the angular
coordinates on a $(\td+1)$-dimensional unit sphere, whose metric is
$d\Omega_{\td+1}^2$. Since we are interested in static
spherically-symmetric solutions, all the functions appearing in the metric
as well as dilaton $\phi$ are assumed to depend only on the radius $r$ of
the transverse dimensions.

If the resulting metric has null isometry, say, in the direction $y_1$,
we can incorporate the boost charge by a well-defined step~\cite{G,CT}.
Since this is quite straightforward, we simply concentrate on the diagonal
metric~(\ref{met}).

For background field strengths, we take the most general ones consistent
with the field equations and Bianchi identities.
The background for an electrically charged $q_{A}$-brane is given by
\bea
F_{0 \a_1 \cdots {\alpha}_{q_{A}} r} = \e_{\a_1 \cdots {\alpha}_{q_{A}}} E',
\hs{3} (n_A = q_A+2),
\label{ele}
\ena
where $\a_1, \cdots, {\alpha}_{q_{A}}$ stand for the compact dimensions.
Here and in what follows, a prime denotes a derivative with respect to $r$.

The magnetic case is given by
\bea
F^{\a_{q_{A}+1} \cdots \a_p a_1 \cdots a_{\td+1}} = \frac{1}{\sqrt{-g}}
e^{-a_{A}\phi} \e^{\a_{q_{A}+1} \cdots \a_p a_1 \cdots a_{\td+1}r}
{\tilde E}',
\hs{3} (n_A =D-q_{A}-2),
\label{mag}
\ena
where $a_1, \cdots, a_{\td+1}$ denote the angular coordinates of the
$(\td+1)$-sphere.
The functions $E$ and $\tilde E$ are again assumed to depend only on $r$.

The electric background (\ref{ele}) trivially satisfies the Bianchi
identities but the field equations are nontrivial. On the other hand, the
field equations are trivial but the Bianchi identities are nontrivial
for the magnetic background (\ref{mag}).

In the above metric~\p{met}, the function $f$ is introduced to describe
the type 2 non-extreme solutions. Here we also define nonvanishing function
\bea
\sum_{\a=0}^p u_\a + \td B=\ln X,
\label{ans}
\ena
to describe type 2 non-extreme extension.
In ref.~\cite{NO1}, the field equations~(\ref{fe}) were solved with the
simplifying ansatz that the combination~\p{ans} vanishes.
Although this was the only assumption there, we show here that it is
not mandatory and that the field equations~(\ref{fe}) can be solved
in a wider context without such ansatz.\footnote{This deformation was also
considered in ref.~\cite{ZZ} for a single brane and in \cite{OPS} for
intersecting branes in pp-wave spacetime.
There the function $f(r)$ in the metric was put to 1.}

In order to solve the field equations~(\ref{fe}), we need the Ricci tensors
for our metric~(\ref{met}). The non-zero components are
\bea
R_{00} &=& e^{2(u_0 - B)} f^{2}\Bigg[ \left(u_0+{\shalf}\ln f\right)''
+ \left(\frac{f'}{f}+\frac{X'}{X}+ \frac{\tilde d+1}{r}\right)
\left(u_0+{\shalf}\ln f\right)'\Bigg], \nn
R_{\a\b} &=& -e^{2(u_\a - B)} f\Bigg[ u_\a'' + \left(\frac{f'}{f}
+\frac{X'}{X}+ \frac{\tilde d+1}{r}\right)
u_\a'\Bigg] \d_{\a\b}, \hs{3} (\a,\b=1,\cdots,p), \nn
R_{rr} &=& -\left(B+{\shalf}\ln f+\ln X\right)'' - \sum_{\a=0}^{p}
\left(u_\a'\right)^2 - {\tilde d}\left(B'\right)^2
+ \left(\frac{X'}{X}-\frac{\tilde d+1}{r}\right)B' \nn
&& -\frac{f'}{2f}\left(2u_0' + \frac{f'}{f}+ \frac{X'}{X}
+\frac{\tilde d+1}{r} \right), \nn
R_{ab} &=& -f \Bigg[ (B+\ln r)'' +\left(\frac{f'}{f}+\frac{X'}{X}
+ \frac{\tilde d+1}{r} \right)(B+\ln r)'
\Bigg] g_{ab} + \frac{\tilde d}{r^2} g_{ab},
\label{ricci}
\ena
where $g_{ab}$ is the metric for ${(\tilde d+1)}$-sphere of radius $r$.

For both cases of electric~(\ref{ele}) and magnetic~(\ref{mag})
backgrounds, we find that the field equations~(\ref{fe}) are cast into
\begin{equation}
\left( u_0 + \shalf\ln f \right)'' + \left( \frac{f'}{f} +\frac{X'}{X}
+ \frac{\td+1}{r} \right) \left( u_0 + \shalf\ln f \right)'
= \frac{1}{f} \sum_{A} \frac{D-q_A-3}{2(D-2)} S_A ({E_A}')^2,
\label{1}
\end{equation}
\begin{equation}
{u_\a}'' + \left( \frac{f'}{f} +\frac{X'}{X}+ \frac{\td+1}{r} \right) {u_\a}'
= \frac{1}{f} \sum_{A} \frac{\d_A^{(\a)}}{2(D-2)} S_A ({E_A}')^2,
\hs{3} (\a=1,\cdots,p),
\label{2}
\end{equation}
\bea
&& \left(B+{\shalf}\ln f+\ln X\right)'' + \sum_{\a=0}^{p} \left(u_\a'\right)^2
+ {\tilde d}\left(B'\right)^2- \left(\frac{X'}{X}-\frac{\tilde d+1}{r}\right)
B' \nn
&& +\frac{f'}{2f}\left(2u_0' + \frac{f'}{f}+ \frac{X'}{X}+\frac{\tilde d+1}{r}
\right)
= -\shalf (\phi')^2 + \frac{1}{f} \sum_{A} \frac{D-q_A-3}{2(D-2)}S_A
({E_A}')^2,
\label{3}
\ena
\begin{equation}
f \left[ (B + \ln r)'' + \left( \frac{f'}{f} +\frac{X'}{X}+ \frac{\td+1}{r}
\right) \left( B + \ln r \right)' \right] - \frac{\td}{r^2}
= - \sum_{A} \frac{q_A+1}{2(D-2)} S_A ({E_A}')^2,
\label{4}
\end{equation}
\begin{equation}
r^{-(\td+1)}X^{-1} \left( r^{\td+1} f X \phi' \right)' = - \sum_{A}
\frac{\e_A a_A}{2} S_A ({E_A}')^2,
\label{5}
\end{equation}
\begin{equation}
\left( r^{\td+1} X S_A {E_A}' \right)' = 0,
\label{6}
\end{equation}
where $A$ denotes the kinds of $q_A$-branes and we have defined
\bea
S_A \equiv \exp \left( \e_A a_A \phi - 2 \sum_{\a \in q_A} u_\a \right),
\label{7}
\ena
and
\bea
\d_A^{(\a)} = \left\{ \begin{array}{l}
D-q_A-3 \\
-(q_A+1)
\end{array}
\right.
\hs{5}
{\rm for} \hs{3}
\left\{
\begin{array}{l}
y_\a \hs{3} {\rm belonging \hs{2} to} \hs{2} q_A{\rm -brane \hs{2} and}
\hs{2} \a=0 \\
{\rm otherwise}
\end{array},
\right.
\label{8}
\ena
and $\e_A= +1 (-1)$ corresponds to electric (magnetic) backgrounds.
For magnetic case we have dropped the tilde from $E_A(r)$. Equations
(\ref{1}), (\ref{2}), (\ref{3}) and (\ref{4}) are the $00, \a\a, rr$ and
$ab$ (angular coordinates) components of the Einstein equation in
eq.~(\ref{fe}), respectively. The last one is the field equation for
the field strengths of the electric backgrounds and/or Bianchi identity
for the magnetic ones.

{}From eq.~(\ref{6}), one finds
\bea
r^{\td+1} XS_A {E_A}' = c_A,
\label{const}
\ena
where $c_A$ is a constant.
With the help of eq.~(\ref{const}), eq.~(\ref{1}) can be rewritten as
\bea
\left[ r^{\td +1} fX \left( u_0 + \shalf\ln f \right)' \right]'
= \sum_{A} \frac{D-q_A-3}{2(D-2)} c_A E_A',
\ena
which can be integrated to give
\bea
f X\left( u_0 + \shalf\ln f \right)' = \sum_{A} \frac{D-q_A-3}{2(D-2)}
 c_A \frac{E_A}{r^{\td+1}} + \frac{c_0 \td}{r^{\td+1}},
\label{ffint}
\ena
where $c_0$ is an integration constant.
Similarly, we find that eqs.~(\ref{2}) and (\ref{5}) give
\bea
fX {u_\a}' &=& \sum_{A} \frac{\d_{A}^{(\a)}}{2(D-2)}
 c_A \frac{E_A}{r^{\td+1}} + \frac{c_\a \td}{r^{\td+1}}, \hs{3}
 (\a=1,\cdots,p), \nn
fX \phi' &=& - \sum_{A} \frac{\e_A a_A}{2} c_A \frac{E_A}{r^{\td+1}}
 + \frac{c_\phi \td}{r^{\td+1}},
\label{fint}
\ena
where $c_\a$ ($\a=1,\cdots,p$) and $c_\phi$ are again integration
constants. We find from eq.~(\ref{4}) the result
\begin{equation}
fX \left( B + \ln r \right)' - \frac{\td}{r^{\td+1}}\int r^{\td-1}X{\rm d}r
= - \sum_{A} \frac{q_A+1}{2(D-2)}c_A \frac{E_A}{r^{\td+1}}
+ \frac{c_b \td}{r^{\td+1}},
\label{B1}
\end{equation}
where $c_b$ is another integration constant. These equations involve
an unknown function $X(r)$ and appear intractable. However, $X(r)$ is not an
independent variable but is given by~\p{ans}. We now show that $X(r)$ and
$f(r)$ can be determined from a constraint and that other functions
$u_\a(r)$ ($\a=0,\cdots,p$), $\phi(r)$ and $B(r)$ can then be solved
consistently together with the electric (magnetic) background $E_A(r)$.

Using the definition of $X(r)$, we can combine eqs.~(\ref{1}),
(\ref{2}) and (\ref{4}) appropriately to derive the constraint
satisfied by $X(r)$ and $f(r)$:
\begin{equation}
\frac{X''}{X}+\left(\frac{3}{2}\,\frac{f'}{f}+\frac{2\td+1}{r}\right)
\frac{X'}{X}
+ \frac{1}{2}\,\frac{f''}{f}+\frac{(3\td+1)}{2r}\,\frac{f'}{f}
+\frac{(f-1)}{f}\,\frac{{\td}^2}{r^2}=0.
\label{Xf}
\end{equation}
Note that there are terms independent of $X$. Since $X$ and $f$ can be
regarded as independent functions, it is natural to set the $X$-independent
part to 0: \footnote{There is the freedom of reparametrization of the
coordinates in the metric~\p{met}. This $f(r)$ corresponds to a
choice of gauge without any loss of generality. This choice is useful
to make the interpolation between the solutions of type 1 and 2 manifest.}
\begin{equation}
f'' + \frac{(3\td+1)}{r} f' + 2(f-1)\frac{{\td}^2}{r^2}=0.
\label{df}
\end{equation}
Solving this second order differential equation gives
$f(r)=(1-\frac{\mu_1}{r^\td})(1-\frac{\mu_2}{r^\td})$ with two integration
constants $\mu_1$ and $\mu_2$. It turns out, however, that the parameter
$\mu_2$ can be absorbed if we redefine the coordinate as
$\tilde r^\td =r^\td -\mu_2$ and $\mu_1$ is shifted by $\mu_2$.\footnote{
This shift is not a symmetry of the system, and it may appear strange that
$\mu_2$ can be absorbed by this. We have actually solved all the field
equations keeping $\mu_1$ and $\mu_2$ and found that the parameter $\mu_2$
could be eliminated by this shift after cancellation of various factors.
For example, if we put $f(r)=f_1(r)f_2(r)$ into eq.~\p{Xf} with
$f_i(r)=1-\frac{\mu_i}{r^\td} \;(i=1,2)$, we get
$X=1-(\nu-1)\frac{(f_1^{1/2}-f_2^{1/2})^2}{2\sqrt{f_1 f_2}}$. After the shift,
we find $f_1(r)=(1-\frac{\mu_1-\mu_2}{\tilde r^\td})f_2(r)$, and $f_2(r)$
drops out of $X(r)$, giving eq.~\p{X}.
The same observation is also made for the solutions
found in ref.~\cite{NS}.}
So we can simply put $\mu_2=0$ without loss of generality and set
\bea
f(r) = 1-\frac{\mu}{r^{\td}},
\label{f}
\ena
which characterizes the type 2 non-extreme extension.
Using eq.~(\ref{f}) in eq.~(\ref{Xf}), we find
\begin{equation}
X(r)=1- (\nu-1)\frac{\left(f^{1/2}-1\right)^2}{2 \sqrt{f}},
\label{X}
\end{equation}
where $\nu$ is yet another integration constant. The choice $\nu=1$ reduces the
solution to type 2 non-extreme case. Thus this parameter $\nu$ introduces
another direction of non-extremality. Note that the function $X$ should
contain in general two arbitrary constants, one of which is eliminated
by the requirement of asymptotic flatness: $u_\a(r)$ $(\a=0,\cdots,p)$,
$\phi(r)$, $B(r) \to 0$ for $r \to \infty$ requires $X(r) \to 1$.

Using eqs.~\p{4}, \p{ffint}, \p{fint}, \p{B1} and \p{X} in \p{3} yields
\bea
&&\left( \sum_{A} \frac{D-q_A-3}{2(D-2)} c_A \frac{E_A}{r^{\td+1}}- \shalf f'X
 + \frac{c_0 \td}{r^{\td+1}}\right)^2
+ \sum_{\a=1}^p \left( \sum_{A} \frac{\d_{A}^{(\a)}}{2(D-2)} c_A
 \frac{E_A}{r^{\td+1}} + \frac{c_\a \td}{r^{\td+1}} \right)^2 \nn
&&+ \td \left( - \sum_{A} \frac{q_A+1}{2(D-2)} c_A \frac{E_A}{r^{\td+1}}
+\frac{1}{r}\left[\nu-(\nu-1)f^{1/2}-fX \right]
 + \frac{c_b \td}{r^{\td+1}}\right)^2 \nn
&& + \shalf \left( - \sum_{A} \frac{\e_A a_A}{2} c_A \frac{E_A}{r^{\td+1}}
 + \frac{c_\phi \td}{r^{\td+1}} \right)^2
+f'X\left( \sum_{A} \frac{D-q_A-3}{2(D-2)} c_A \frac{E_A}{r^{\td+1}}
- \shalf f'X + \frac{c_0 \td}{r^{\td+1}}\right)\nn
&&-fX\left(\frac{f'}{f}+2\frac{X'}{X}\right)
\left( - \sum_{A} \frac{q_A+1}{2(D-2)} c_A \frac{E_A}{r^{\td+1}}
+\frac{1}{r}\left[\nu-(\nu-1)f^{1/2}-fX \right]
 + \frac{c_b \td}{r^{\td+1}}\right)\nn
&&+fX^2\left[\frac{f''}{2}+f\left(\frac{X'}{X}\right)'+\frac{\td-1}{2r}f'
+\left(\frac{f'}{2}-\frac{f}{r}\right)\frac{X'}{X}-(f-1)\frac{\td}{r^2}
\right] \nn
&&=\shalf fX\sum_{A} \frac{c_A}{r^{\td+1}} {E_A}'.
\label{sint}
\ena
This equation must be valid for functions $E_A$ of $r$.

With the help of eqs.~(\ref{f}) and~(\ref{X}), the $E_A$-independent
part of eq.~(\ref{sint}) yields a constraint condition among the constants
introduced above:
\begin{equation}
\sum_{\a=0}^pc_{\a}^2+\td c_b^2 + \shalf c_{\phi}^2
-\frac{\td+1}{2\td}\left(\nu-\shalf\right){\mu}^2=0,
\label{constant}
\end{equation}
where we have redefined $c_b$ by a constant shift
($c_b \to c_b-\frac{\mu\nu}{2\td}$).
The $E_A$-dependent part of eq.~(\ref{sint}), on the other hand, can be
rewritten as
\bea
\sum_{A,B} \left[ M_{AB} \,\frac{c_A}{2} + \left(r^{\td+1}fX
\left( \frac{1}{E_A} \right)' +\frac{\tilde c_A}{E_A} \right) \d_{AB}
\right] \frac{c_B}{2}\, \frac{E_A E_B}{r^{2\td+2}}
=0,
\label{tint}
\ena
where
\bea
M_{AB} = \sum_{\a=0}^p \frac{\d_{A}^{(\a)}\d_{B}^{(\a)}}{(D-2)^2}
 + \td \frac{(q_A+1)(q_B+1)}{(D-2)^2} + \shalf \e_A a_A \e_B a_B,
\label{cond1}
\ena
and
\bea
\tilde c_A = 2\td \sum_{\a=0}^{p}
\frac{\d^{(\a)}_A}{D-2}\,c_\a - 2\td^2 c_b\frac{q_A + 1}{D-2}
- \tilde d\e_A a_A c_\phi.
\label{tca}
\ena
Note that for $\nu <\frac{1}{2}$, eq.~\p{constant} tells us that
$c_\a=c_b=c_\phi=\mu=0$, and this does not give nontrivial solution.
The same is true for $\nu=\frac{1}{2}$. Hence we restrict ourselves to
$\nu >\frac{1}{2}$. Since $M_{AB}$ is constant, eq.~(\ref{tint}) cannot be
satisfied for arbitrary functions ${E_A}$ of $r$ unless the second term
inside the square bracket is a constant. Substituting eqs.~(\ref{f}) and
(\ref{X}) into this differential equation, one obtains the solution
\bea
E_A(r) = \frac{N_A }{1 - \b_A(1 - g^{-\a_A})},
\label{E}
\ena
where $\b_A$ and $N_A$ are integration constants, and
\bea
g(r)=\left| \frac{f^{1/2} - \rho}{\rho f^{1/2} - 1}
\right|,~~
\a_A = \frac{2}{\td \sqrt{2\nu-1}\; \mu}\:\tilde c_A,
\label{i}
\ena
where parameter $\rho$ is defined as
\bea
\rho \equiv \frac{\nu-1}{\nu+\sqrt{2\nu-1}}.
\label{i2}
\ena

Equation~(\ref{tint}) has two implications if we take
independent functions for the background fields $E_A(r)$. In this case,
first putting $A=B$ in eq.~(\ref{tint}), we learn that
\bea
\frac{c_A}{2} = \frac{\tilde c_A(\b_A-1)}{N_A M_{AA}}
\equiv \frac{\tilde c_A(\b_A-1)}{N_A}\frac{D-2}{\Delta_A},
\label{can}
\ena
where $\Delta_A$ is given in
\bea
\Delta_A = (q_A + 1) (D-q_A-3) + \shalf a_A^2 (D-2).
\label{delta}
\ena
By use of eqs.~(\ref{f}), (\ref{X}), (\ref{E})--(\ref{delta}), we integrate
eqs.~(\ref{ffint})--(\ref{B1}) to obtain the results
\bea
u_0(r) &=&  -\sum_{A} \frac{D-q_A-3} {\Delta_A} \ln H_A
+\frac{2c_0}{\sqrt{2\nu-1}\; \mu} \ln g-\shalf \ln f, \nn
u_\a(r) &=& -\sum_{A} \frac{\d^{(\a)}_A} {\Delta_A} \ln H_A
+\frac{2c_\a}{\sqrt{2\nu-1}\; \mu} \ln g, \hs{3} (\a=1,\cdots,p), \nn
\phi(r) &=& \sum_{A} \e_A a_A  \frac{D-2}{\Delta_A}  \ln H_A
+\frac{2c_\phi}{\sqrt{2\nu-1}\; \mu} \ln g, \nn
B(r) &=&  \sum_{A} \frac{q_A+1}{\Delta_A} \ln H_A
+ \frac{2c_b}{\sqrt{2\nu-1}\; \mu} \ln g+\frac{1}{\tilde d}
\left(\frac{1}{2}\ln f + \ln X\right),
\label{warpn}
\ena
where $H_A(r)$ is given by
\bea
H_A(r) =N_A E_A^{-1} g^{\a_A}
= \left[1-\b_A (1-g^{-\a_A})\right]g^{\a_A},
\label{han}
\ena
and the integration constants are fixed by the requirement that
the metrics approach to $1$ asymptotically.

Using eq.~(\ref{warpn}), one can write down the expression for $S_A(r)$ as
\bea
S_A(r)= N_A^2 E^{-2}_A fg^{\a_A}.
\ena
Now, using eqs.~(\ref{const}) and (\ref{can}), we can determine
the normalization constant $N_A$ as
\bea
N_A^2 = \frac{2(\b_A-1)}{\b_A}\frac{(D-2)}{\Delta_A}.
\ena
We also have
\bea
\sum_{\a=0}^{p} c_\a + \tilde d c_b =0,
\label{condnn}
\ena
from the relation~(\ref{ans}). By use of this relation, $\tilde c_A$ in
eq.~(\ref{tca}) can also be written as
\bea
\tilde c_A = \tilde d \left(2 \sum_{\a \in q_A} c_\a - \e_A a_A c_\phi \right).
\label{cons}
\ena

Our metric and background fields are thus finally given by, after
putting all the warp factors etc. that we get by solving the
Einstein equations,
\bea
ds_D^2 &=& \prod_A {H}_A^{2 \frac{q_A+1}{\Delta_A}} \Bigg[ - \prod_A
{H}_A^{- 2 \frac{D-2}{\Delta_A}} g^{4 c_0/(\sqrt{2\nu-1}\; \mu)}dt^2
+ \sum_{\a=1}^{p} \prod_A H_A^{- 2 \frac{\c_A^{(\a)}}{\Delta_A}}
g^{4 c_\a/(\sqrt{2\nu-1}\; \mu)}dy_\a^2
\nn
&&+(fX^2)^{1/\td}g^{4 c_b/(\sqrt{2\nu-1}\; \mu)}\left(f^{-1} dr^2
+ r^2 d\Omega_{\td+1}^2\right) \Bigg], \nn
E_A(r) &=& \pm \sqrt{2 \frac{\b_A-1}{\b_A}\frac{D-2}{\Delta_A}}\:
{H}^{-1}_A g^{\a_A},
\label{metn}
\ena
where we have defined
\bea
\c_A^{(\a)} = \left\{ \begin{array}{l}
D-2 \\
0
\end{array}
\right.
\hs{5}
{\rm for} \hs{3}
\left\{
\begin{array}{l}
y_\a \hs{3} {\rm belonging \hs{2} to} \hs{2} q_A{\rm -brane} \\
{\rm otherwise}
\end{array}.
\right.
\ena

The second condition following from eq.~(\ref{tint}) is
$M_{AB}=0$ for $A \neq B$. As shown in ref.~\cite{NO1}, this leads to
the intersection rules for two branes. If $q_A$-brane and $q_B$-brane
intersect over ${\bar q}\; (\leq q_A, q_B)$ dimensions, this gives
\bea
{\bar q} = \frac{(q_A+1)(q_B+1)}{D-2}-1 - \shalf \e_A a_A \e_B a_B.
\label{int}
\ena
{}For eleven-dimensional supergravity, we have electric 2-branes, magnetic
5-branes and no dilaton $a_A=0$. The rule (\ref{int}) tells us that
2-brane can intersect with 2-brane on a point $(\bar q=0)$ and with
5-brane over a string $(\bar q=1)$, and 5-brane can intersect with
5-brane over 3-brane $(\bar q=3)$, in agreement with
refs.~\cite{PT,T}.

The solutions~\p{metn} are the general intersecting branes which interpolate
non-extreme solutions of type 1 and 2.
As mentioned before, for $\nu=1$, we have $X=1, g=f^{1/2}$ and the above
solutions give generalized non-extreme solutions of type 2 with
$p+m+2$ parameters $c_\a (\a=0,\cdots,p), c_b, c_\phi,\b_A (A=1,\cdots, m)$
and $\mu$ restricted by 2 constraints~\p{constant} and \p{condnn}.
If we further choose $c_0=\frac{\mu}{2}, c_b=-\frac{\mu}{2 \td},
c_\a=c_\phi=0, (\a=1,\cdots, p)$, they reduce to the known solutions
(see for example \cite{NO1}).

It appears that they no longer give non-extreme solutions of type 1
if we put $\mu=0$ since then the non-extreme function $X$ in \p{X}
becomes 1. However, we can manage to derive such solutions as follows.
Consider the limit sending $\mu$ to zero. If we keep the combination
\bea
\frac{\nu-1}{8} \mu^2 \equiv r_0^{2\td},
\label{limit1}
\ena
finite, we get nontrivial functions
\bea
X(r)=1-\left(\frac{r_0}{r}\right)^{2\td};\hs{10}
g(r) = \frac{1-(r_0/r)^{\tilde d}}{1+(r_0/r)^{\tilde d}}.
\label{spx}
\ena
It is then easy to see that the solutions reproduce the non-extreme
ones of type 1 discussed in ref.~\cite{OPS}.

It would be instructive to explicitly give the single brane case.
The metric is
\bea
ds_D^2 &=& {H}^{ \frac{2(p+1)}{\Delta}} \Bigg[
{H}^{- 2 \frac{D-2}{\Delta}}\Bigg(- g^{4c_0/(\sqrt{2\nu-1}\; \mu)}dt^2
+ g^{4c_u/(\sqrt{2\nu-1}\; \mu)}\sum_{\a=1}^{p}dy_{\a}^2\Bigg)
\nn
&&+(fX^2)^{{1}/{\tilde d}}g^{4c_b/(\sqrt{2\nu-1}\; \mu)}\left(f^{-1} dr^2
+ r^2 d\Omega_{\td+1}^2\right) \Bigg],
\label{sb}
\ena
where we have set $c_1=c_2=\cdots=c_{p} \equiv c_u,
{\tilde c} = {\tilde d} [2c_0+2p c_u-\e ac_{\phi}]$, and the other quantities
$\Delta, H(r)$ and $\a$ are given in \p{delta}, \p{han} and \p{i}
with the subscript $A$ (which is irrelevant for a single brane) removed
and $q$ replaced by $p$, respectively.
There are five independent parameters in the above single brane
metric. Namely we have seven integration constants $c_0, c_u, c_b, c_{\phi},
\beta, \nu$ and $\mu$ restricted by the two
constraints from eqs.~\p{constant} and \p{condnn}.

Our single brane solution includes that of ref.~\cite{ZZ} as a special
case which is a four parameter solution. If we consider the limit
$\mu \to 0$ keeping eq.~\p{limit1} finite,
our solution~\p{sb} reduces to a single brane case with
$X(r)$ and $g(r)$ in eq.~\p{spx}, and
$\a = \frac{1}{2{\tilde d}r_0^{\tilde d}}\,{\tilde c}$,
with constraints
\begin{equation}
c_{0}^2+p c_{u}^2+{\tilde d}c_b^2+\frac{1}{2}c_{\phi}^2
-4\frac{{\tilde d}+1}{\tilde d}r_{0}^{2{\tilde d}}=0,
\end{equation}
and \p{condnn}.
This solution contains four independent parameters ($c_0, c_u, c_b, c_{\phi},
\beta$ and $r_0$ restricted by the two constraints).
It is easy to transform our solution to the complete solution of \cite{ZZ}
with redefinition of parameters.

To summarize, we have given very general intersecting brane solutions
without assuming any restriction on the metric such as~\p{x}.
The result is a general class of the brane solutions which interpolate
the non-extreme solutions of type 1 and 2, which are expected to give
further insight into the nonperturbative effects in string and
field theories. The method we use is a simple generalization of the one
in ref.~\cite{NO1}, which can also be applied to time-dependent cases as
well~\cite{NO2}. It is gratifying to find that the method is so useful.


\vspace{10mm}
\noindent
{\bf Acknowledgments}

This work was supported in part by Grants-in-Aid for Scientific Research
Nos.12640270 and 02041. Y.-G. M. also acknowledges the support from
the National Natural Science Foundation of China under grant No.10275052.

\newcommand{\NP}[1]{Nucl.\ Phys.\ B\ {\bf #1}}
\newcommand{\AP}[1]{Ann.\ Phys.\ {\bf #1}}
\newcommand{\PL}[1]{Phys.\ Lett.\ B\ {\bf #1}}
\newcommand{\NC}[1]{Nuovo Cimento {\bf #1}}
\newcommand{\CMP}[1]{Comm.\ Math.\ Phys.\ {\bf #1}}
\newcommand{\PR}[1]{Phys.\ Rev.\ D\ {\bf #1}}
\newcommand{\PRL}[1]{Phys.\ Rev.\ Lett.\ {\bf #1}}
\newcommand{\PRE}[1]{Phys.\ Rep.\ {\bf #1}}
\newcommand{\PTP}[1]{Prog.\ Theor.\ Phys.\ {\bf #1}}
\newcommand{\PTPS}[1]{Prog.\ Theor.\ Phys.\ Suppl.\ {\bf #1}}
\newcommand{\MPL}[1]{Mod.\ Phys.\ Lett.\ {\bf #1}}
\newcommand{\IJMP}[1]{Int.\ Jour.\ Mod.\ Phys.\ {\bf #1}}
\newcommand{\JP}[1]{Jour.\ Phys.\ {\bf #1}}

\end{document}